# Past Annual Variations of the Karst Denudation Rates.

**Shopov Y. (1), Stoykova D. (1), L.T.Tsankov (1), E. Marinova (1), Sauro U. (2), Borsato A. (3), Cucchi F. (4), Forti P. (5), Piccini L. (6), D. C. Ford (7), Chas.J. Yonge (8)**

(1) Faculty of Physics, University of Sofia, 1164 Sofia, Bulgaria, E-mail: YYShopov@Phys.Uni-Sofia.BG
(2) Dipartimento di Geografia, Università di Padova, Via del Santo 26, 35123 Padova, Italy,
(3) Museo Tridentino di Scienze Naturali, Via Calepina 14, 38100 Trento, Italy,
(4) Dipartimento di Scienze Geologiche, Ambientali e Marine, Universita' di Trieste, Via Weiss, 2, 34127 Trieste, Italy,
(5) Dipartimento di Scienze della Terra e Geologico Ambientali, Via Zamboni 67, 40127, Bologna, Italy,
(6) Dipartimento di Scienze della Terra, Universita' di Firenze, Via La Pira 4, 50121 Firenze, Italy,
(7) Geography Dept., McMaster University, Hamilton, Ontario, L8S 1K4, Canada,
(8) Dept. of Physics, University of Calgary, Calgary, Alberta, Canada

**Abstract**

We used the quantitative theory of solubility of karst rocks of Shopov et. al, (1989, 1991a) in dependence of the temperature and other thermodynamic parameters to make reconstructions of past carbonate denudation rates. This theory produced equations assessing the carbonate denudation rates in dependence on the temperature or on the precipitation. We used an estimate of the averaged denudation rate in the region based on integrated data of the carbonate hardness or the water from springs, rivers, cave pools and dripping water and average precipitation rate (of 470 mm/yr) from meteorological data for Kananaskis karst region, Alberta, Canada. Obtained denudation rate is 14 mm/kyr or 38 t/km$^2$ per year. We used this estimate as starting point and substituted our proxy records of the annual temperature and the annual precipitation in the equations of dependence of karst denudation rate on precipitation and temperature. This way we reconstructed variations of the annual karst denudation rate for the last 280 years in dependence on the annual precipitation and for the last 1250 years in dependence on the temperature. Both reconstructions are made for equilibrium conditions and do not take into account variations of the evapotranspiration, but they produce quite reasonable estimate of the variations of carbonate denudation, which is within observed variation of 8- 20 mm/kyr (86% variation). Precipitation dependence of carbonate denudation produces 79 % variation in the denudation rate in result of the reconstructed variation of 300 mm/yr from the driest to the wettest year during the last 280 years. Temperature dependence of carbonate denudation due to temperature dependence of solubility of the carbonate dioxide produce only 9.3 % variation in the denudation rate in result of the reconstructed variation of 4.7 deg. C during the last 1250 years, so it is negligible in respect of the precipitation dependence.

We measured a very high- resolution luminescence record covering the last 2028 ±100 years from Savi Cave, Trieste, Italy, which allows precise determination of growth rate of the stalagmite. It consists of 40106 data points compiled of 16 overlapping scans. Its time step varies from 15.6 days to 19.9 days. We made a reconstruction of the annual growth rate variations for the last 2028 years, which represents annual precipitation for the region of the cave. It allows us to estimate the range of the annual variations of the karst denudation in North Italy.

We found that the strongest cycle of the annual rainfall in the region of Trieste, Italy is with duration of about 300 years. Several other cycles with duration of 160, 130, 68, 38, 30.2, 18.4, 9.4, 6.8 and 5.8 years exist in the precipitation there. They should produce variations with the same duration in the karst denudation rates in the region.

### Introduction

Calcite speleothems usually display luminescence produced by calcium salts of humic and fulvic acids derived from soils above the cave (Shopov, 1989; White and Brennan; 1989). These acids are released by the roots of living plants and by the decomposition of dead vegetative matter. Root release is modulated by the visible solar radiation via photosynthesis, while rates of decomposition depend exponentially on soil temperature. Soil temperature depends mainly on solar infrared and visible radiation in the case that the cave is covered only by grass or upon air temperature in the case that the cave is covered by forest or bush (Shopov et al., 1994). In the first case the zonality of luminescence can be used as a proxy of solar insolation (Shopov et al., 1990) and in the second case it can be used as a paleotemperature proxy.

The luminescent index has high resolution as in the case when the step of a record is less than one month the signal contains mainly climatic modulation. But in the case of the step bigger than one year the climatic modulation of signal is in the range of the experimental error and the luminescent record turns to proxy of solar insolation.

Speleothem growth rate variations represent mainly rainfall variations (Shopov et al., 1992, 1994). Speleothem luminescence visualises annual microbanding (Shopov et al. 1991a). We used it to derive proxy records of the annual precipitation at the cave site by measuring the distance between all adjacent annual maxima of the intensity of luminescence. The resultant growth rates correlate with the actual annual precipitation (summed from August to August).

487

## Experimental Part

High-resolution fluorescence records are obtained using LLMZA analysis equipment described in (Shopov, (1987) with excitation wavelength from 200 to 240 nm. Obtained pictures are scanned by precise professional scanner EPSON 1650 with a step of 8 μm. We choose a string with width of 200 μm from scans and transform the string into luminescent curve by a computer program specially made for this purpose. This program integrates pixels contained in a window of 20 x 200 μm that is moving along the string with a step of 10 μm. This digital procedure is equal to the scanning of negatives by a scanning microdensitometer (PDS or Joyce Loebl 6). So this way is obtained a record of distribution of the optical density luminescence (decimal logarithm of the intensity of luminescence) of the speleothem along it's growth axis. It is linearly proportional to the concentration of the luminescent compounds in the calcite. Such records are proxies of the solar radiation or paleotemperature in the past if all these compounds are only organic (Shopov, 1997).

Luminescent records were transformed into luminescent time series by using absolute dating.

## Results and Analyses

We studied a 35 mm long stalagmite from Rats Nest cave (RNC), Alberta, Canada. We measured a stacked 66000- data point luminescent record from Rats Nest cave, Kananaskis karst region, Alberta, Canada. It covers last 1450 yrs with resolution of about 8 days for most of the time span (Shopov, et al., 1998). Paleoclimatic records have been derived from speleothem luminescence by calculation of the average annual intensity of luminescence and measurements of annual growth rate values. Obtained annual records has been calibrated by actual climatic records from near climatic station in Banff, Alberta, located in the same valley, 50 km northern of the cave (Shopov et. al, 1996 a, b). This way we reconstructed annual air temperatures for last 1450 years at the cave site with an estimated error of 0.35 $^0$C, while the error of the direct measurements is 0.1 $^0$C.

Speleothem growth rate variations represent mainly rainfall variations. We obtained a reconstruction of the annual precipitation for the last 280 years at the cave site. The estimated statistical error is 80 mm/year. Annual speleothem growth rate was independent on the intensity of luminescence, on annual temperature and on solar luminosity for the same time span (zero correlation).

Speleothem luminescence visualises annual microbanding we used for relative and absolute dating of speleothems by Autocalibration dating (Shopov et al, 1991 a).

This stalagmite was dated by 14- C and autocalibration dating:

1. The TAMS $^{14}$C dating of this sample produced age of 1450 +/- 150 years (2 sigma) of the base of the stalagmite. The 14-C date is corrected for "dead" carbon, by its measurement in modern speleothem calcite in RNC. **2.** The Autocalibration dating produced better precision of 1450 +/- 80 years.

Both methods produced consistent age of the base of the stalagmite (in the frames of their experimental error)

We used the quantitative theory of solubility of karst rocks of Shopov et. al, (1989, 1991a) in dependence of the temperature and other thermodynamic parameters to make reconstructions of past carbonate denudation rates. This theory produced equations of the dependence of the carbonate denudation rates (**D**) in dependence on the precipitation and on the temperature:

**D= k. R. [CaCO$_3$],**                                                                                                                          (1)

Where **k**- is the ratio between the amount of water passed through the karst rock and the total precipitation, **R** - is precipitation and **[CaCO$_3$]** - concentration of calcium carbonate in the karst waters

**[CaCO$_3$]= [CaCO$_3$]$_0$ . pCO$_2^{1/3}$. e$^{-0.02t}$**                                                                                          **(2)**

Where **pCO$_2$** is the partial pressure of **CO$_2$** inside the bedrock, **[CaCO$_3$]$_0$**- is the solubility of the studied rock at standard thermodynamic conditions (25$^o$ C, 1 atm. **pCO$_2$**) which can be measured it the lab, **t** is the temperature inside the bedrock in $^o$C (Shopov et al., 1998).

Usually **[CaCO$_3$]$_0$ . pCO$_2^{1/3}$** is constant with time. So combining equations 1 and 2 we obtain:

**D= k. R. [CaCO$_3$]$_0$ . pCO$_2^{1/3}$. e$^{-0.02t}$= const. R. e$^{-0.02t}$**                                           **(3)**

If we presume that annual temperature is constant we may study the precipitation dependence of the karst denudation. If we presume that annual precipitation is constant we may study the temperature dependence of the karst denudation. We made this in order to evaluate the importance of the influence of the time variations of both factors on the karst denudation:

We used an estimate of the averaged denudation rate in the region based on integrated data of the carbonate hardness of the water from springs, rivers, cave pools and dripping water and average precipitation rate (of 470 mm/yr) from meteorological data. Obtained denudation rate is 14 mm/kyr or 38 t/km$^2$ per year. We used this estimate as starting point and substituted our proxy record of the annual precipitation (for the last 280 years) in the equation (3), presuming that **t** was constant equal to the average annual temperature of the cave region, to reconstruct the karst denudation variations in dependence on the annual precipitation (fig. 1).

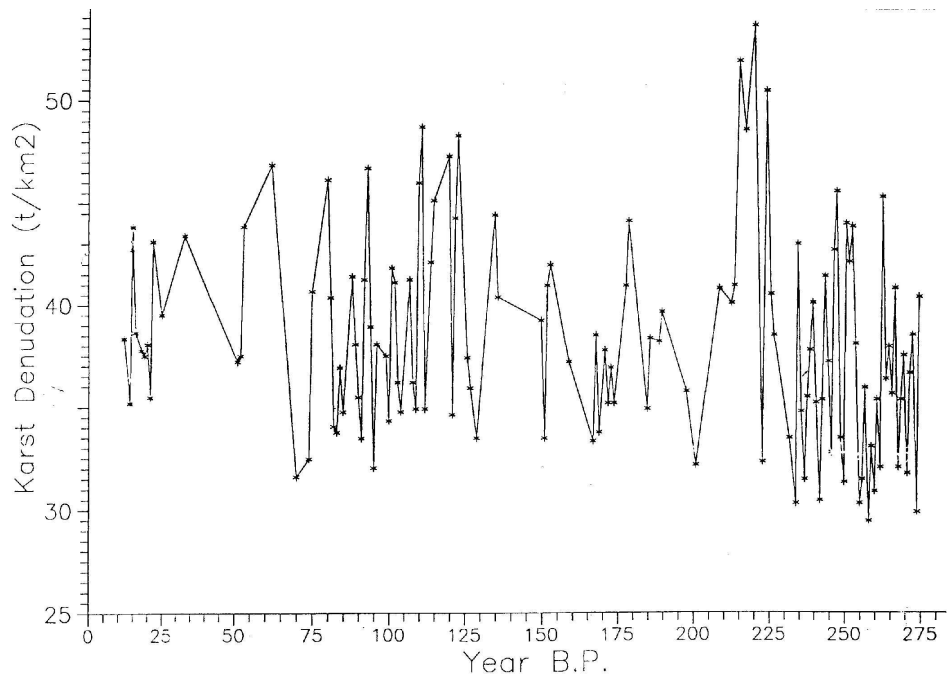

Fig. 1 Reconstruction of variations of carbonate denudation rate in Kananaskis karst region, Alberta, Canada in the last 280 years in dependence on the annual precipitation

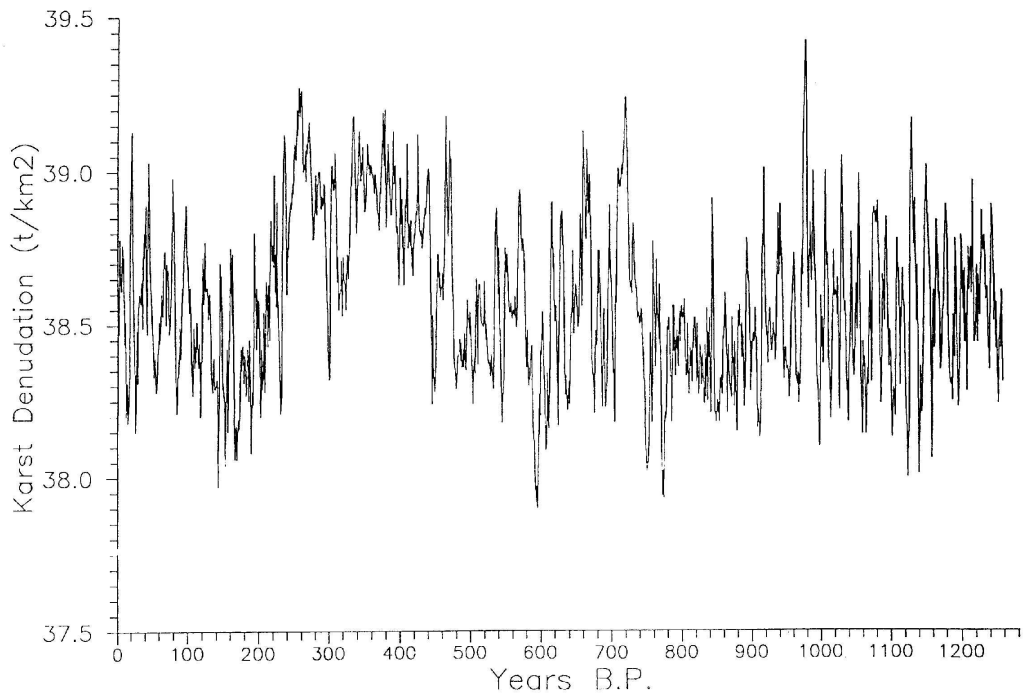

Fig.2. Reconstruction of annual variations of carbonate denudation rate in Kananaskis karst region, Alberta, Canada in the last 1250 years in dependence on the temperature.

We substituted our proxy record of the annual temperature (for the last 1250 years) in the equation (3), presuming that **R** was constant equal to average annual precipitation (of 470 mm/yr) at the cave region, to reconstruct the karst denudation variations in dependence on the temperature (fig. 2). Both reconstructions are made for equilibrium conditions and do not take into account variations of the evapotranspiration, but they produce quite reasonable estimate of the variations of carbonate denudation, which is within observed variation of 8- 20 mm/kyr (86% variation). Precipitation dependence of carbonate denudation produces 79 % variation in the denudation rate in result of the reconstructed variation of 300 mm/yr from the driest to the wettest year during the last 280 years. Temperature dependence of carbonate denudation due to temperature dependence of solubility of the carbonate dioxide produce only 9.3 % variation in the denudation rate in result of the reconstructed variation of 4.7 deg. C during the last 1250 years, so it is negligible in respect of the precipitation dependence.

We studied a sample from Grotta Savi, in the Trieste Karst, NE Italy, which is a 27-cm long, 5-mm thick polished section of an active calcite stalagmite along its growth axis. It has been dated by means of 18 U/Th MC-ICPMS analyses (Borsato et al., 2004). After identification of positions of U/Th dates along the luminescent scans we transformed them into luminescent time series. The whole stalagmite consists of translucent, columnar calcite, and is characterised by visible growth laminae.

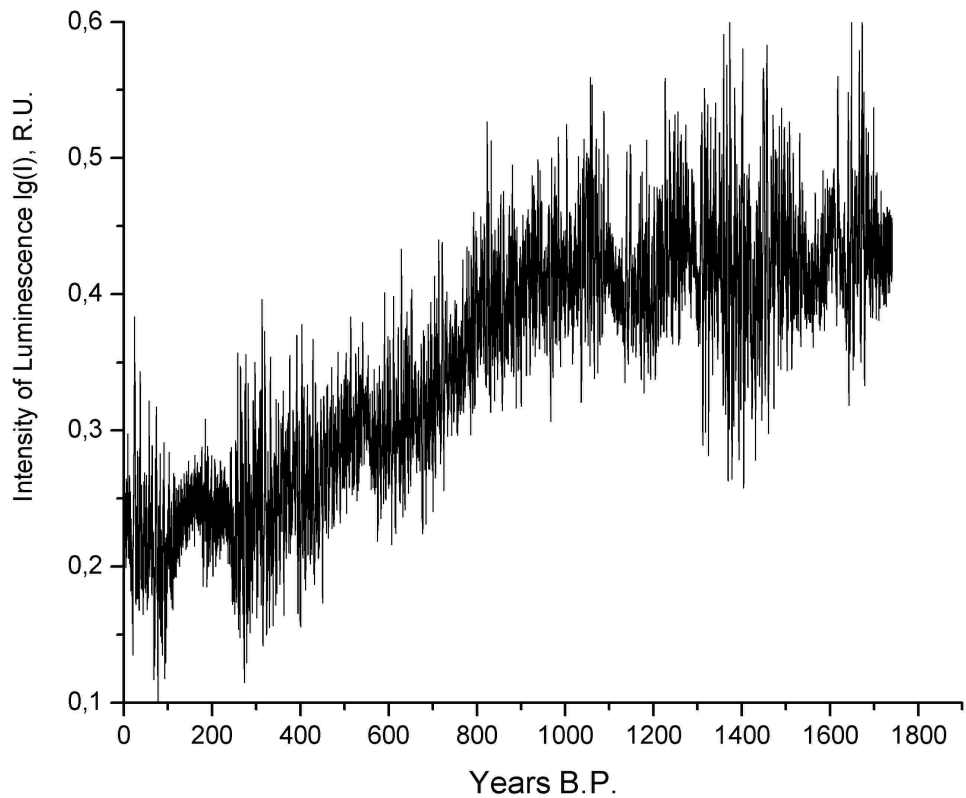

**Fig. 3 The highest resolution composite record of SV1 covers the last 2028±100 years. It consists of 40106**

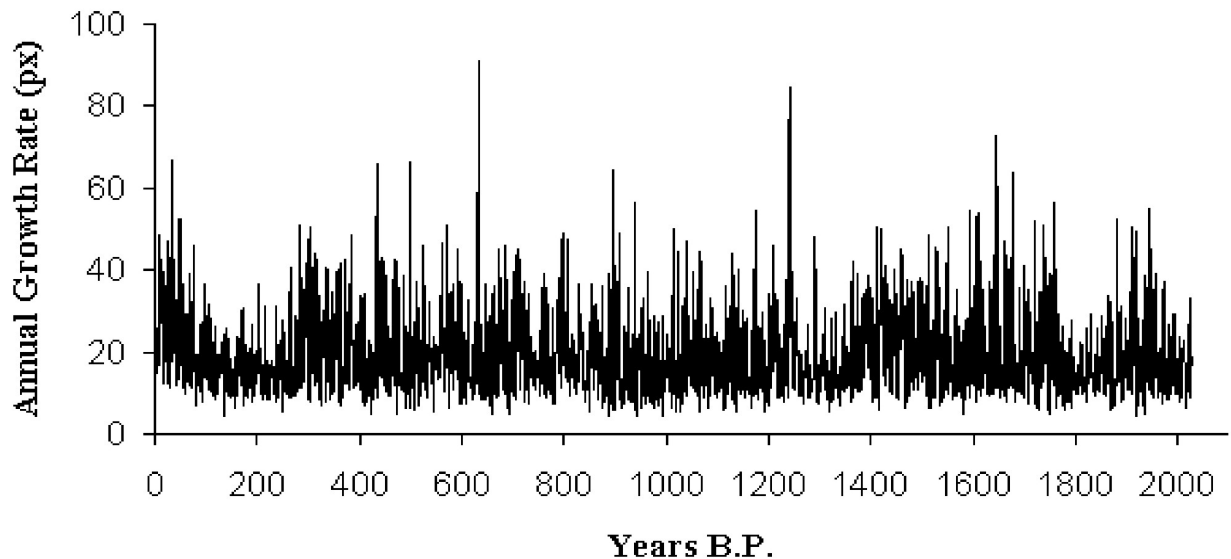

**Fig. 4. The annual growth rate of SV1 stalagmite derived from the record on fig. 3. The mean annual growth rate is 6.36 microns/year and it varies from 2.2 to 45.4 ±0.5 microns/year. This record represents mainly the annual rainfall at the cave site.**

We used a special algorithm, called real-space periodogramme analysis introduced in Shopov et al. (2002) to calculate the intensity of the cycles of the annual precipitation at the cave site. Resulting periodogramme shown on fig. 5 demonstrates that the strongest cycle of the annual rainfall in the region of Trieste, Italy is with duration of about 300 years. Several other cycles with duration of 160, 130, 68, 38, 30.2, 18.4, 9.4, 6.8 and 5.8 years exists in the precipitation there. They should produce variations with the same duration in the karst denudation rates in the region.

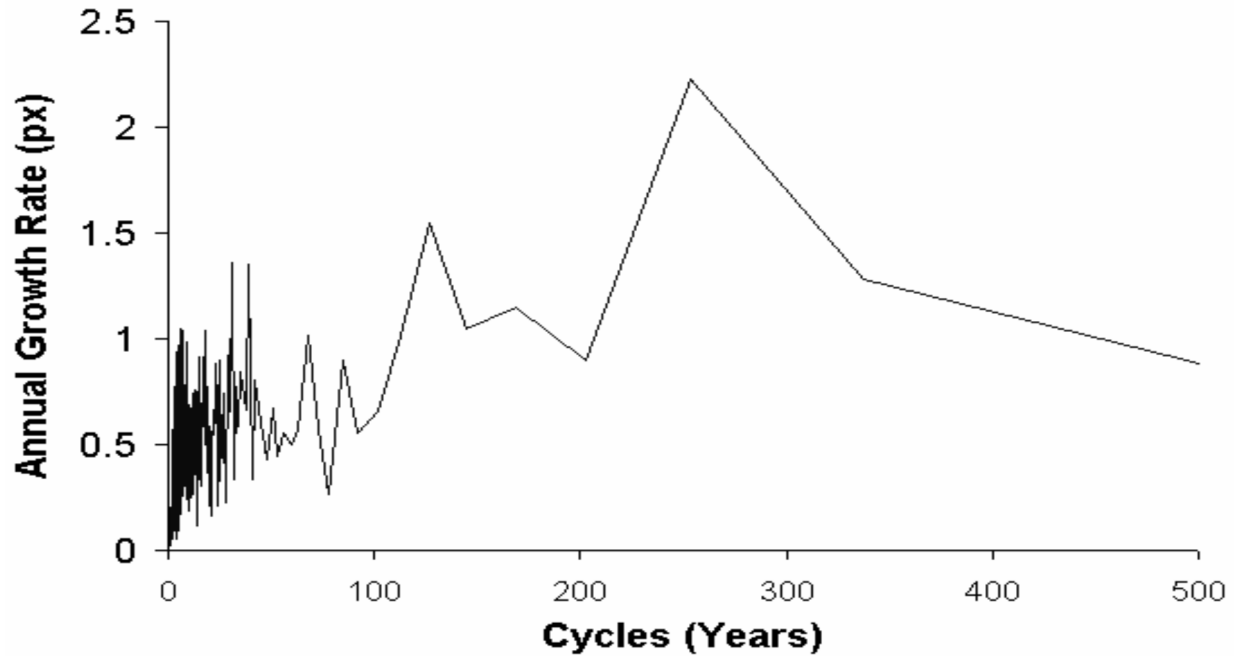

**Fig. 5. A periodogramme of the annual growth rate of SV1 stalagmite from cave Savi. It represents the cycles of the annual rainfall in the region of Trieste, Italy.**

### Conclusion

It is demonstrated that speleothem luminescence proxy records of annual values of the climatic parameters could be used for reconstruction of the carbonate denudation variations for a time span far exceeding all historic records.

It is demonstrated that variation of carbonate denudation due to temperature dependence of solubility of the carbonate dioxide is negligible in respect of the variations due to precipitation changes.

Strongest cycle of the annual rainfall and of the carbonate denudation in the region of Trieste, Italy is with duration of about 300 years. Several other cycles with duration of 160, 130, 68, 38, 30.2, 18.4, 9.4, 6.8 and 5.8 years exists there.

### Acknowledgements

This research has been funded by COFIN 2000 "Ricostruzione dell'evoluzione climatica e ambientale ad alta risoluzione da concrezioni di grotta lungo una traversa N-S in Italia con particolare riferimento all'intervallo Tardiglaciale-attuale", coordinated by U. Sauro and supported by grant NZ 811/ 98 of Bulgarian Science Foundation to Y. Shopov and a NSERC strategic research grant to D.C. Ford.